\documentclass[manyauthors]{fundam}

\publyear{23}
\papernumber{2102}
\volume{185}
\issue{1}

\newcommand {\kw}[1]{\mathit{#1}}
\newcommand {\length}{length}
\newcommand {\concat}{\otimes}
\newcommand {\head}{head}
\newcommand {\tail}{tail}
\newcommand {\modkw}{~mod~}

\newcommand {\emptyrelation}{\phi}

\newcommand {\tabeq}{\hspace*{0.1in}}

\newcommand {\tabcom}{\hspace*{0.25in}}

\newcommand {\nln}{@{}l@{}}
\newlength{\interligne}
\newcommand {\dpreuve}{\dimen123=\linewidth \dimen124=\linewidth
\advance\dimen123 by -20mm \advance\dimen124 by -5mm
\advance\dimen123 by -\mathindent \advance\dimen124 by -\mathindent
\setlength{\interligne}{\baselineskip}
\setlength{\baselineskip}{1.2\baselineskip}
    \begin{tabbing} 
    \hspace*{\mathindent}\= \hspace*{5mm}\= \kill 
    \+ \kill}
\newcommand {\fpreuve}{\end{tabbing}
    \setlength{\baselineskip}{\interligne}}

\newcommand {\dpreuveitem}{\begin{tabbing} 
    \hspace*{5mm}\= \hspace*{15mm}\= \kill}
\newcommand {\fpreuveitem}{\end{tabbing}}

\newcommand {\dspecitem}{\begin{tabbing} 
    \hspace*{5mm}\=\hspace*{5mm}\=\hspace*{5mm}\=
    \hspace*{5mm}\=\hspace*{5mm} \kill}
\newcommand {\fspecitem}{\end{tabbing}}

\def\[{\relax\ifmmode\@badmath\else\begin{trivlist}\item[]\leavevmode
  \hbox to\linewidth\bgroup$ \displaystyle
  \hskip\mathindent\bgroup\fi}
\def\]{\relax\ifmmode \egroup $\hfil \egroup \end{trivlist}\else \@badmath \fi}

\newcommand {\rng}{\mbox{\it rng}}
\newcommand {\dom}{\mbox{\it dom}}

\newcommand {\dtabin}{\begin{tabbing} 
    \hspace*{\mathindent}\= \kill \+ \kill}
\newcommand {\ftabin}{\end{tabbing}}
\newcommand {\dspec}{\begin{tabbing} 
    \hspace*{\mathindent}\= \hspace*{5mm}\=\hspace*{5mm}\=\hspace*{5mm}\=
                            \hspace*{5mm}\=\hspace*{5mm} \kill 
    \+ \kill}

\newcommand {\fspec}{\end{tabbing}}

\newcommand{\begproof}{\vspace{0.01in}\noindent{\bf Proof.  }}

\newcommand {\finproof}{\hspace*{\fill \mbox{{\bf qed}}}\\ \vspace{0.05in}}

\newlength {\longueurtop}
\newcommand {\initlongueurtop}{\setlength{\longueurtop}{\topsep}}
\newcommand {\topzero}{\setlength{\topsep}{0pt}}
\newcommand {\topdefaut}{\setlength{\topsep}{\longueurtop}}
\newcommand {\debuttab}{ \initlongueurtop \topzero \begin{tabbing} }
\newcommand {\fintab}{ \end{tabbing} \topdefaut }

\newcommand {\truekw}{{\bf true}~}

\usepackage{url} 
\usepackage[ruled,lined]{algorithm2e}
\usepackage{graphicx}
\usepackage{tikz}
\usetikzlibrary{automata, positioning, arrows}

\begin{document}

\title{Invariant Relations:\\
A Bridge from Programs to Equations}

\address{Wided Ghardallou, University of Sousse, Tunisia}

\author{Wided Ghardallou\\
University of Sousse \\
Sousse, Tunisia\\
wided.ghardallou{@}gmail.com
\and Hessamaldin Mohammadi\\
New Jersey Institute of Technology \\
Newark, NJ USA\\
hm385@njit.edu
\and Elijah Brick\\
New Jersey Institute of Technology \\
Newark, NJ USA\\
eb275@njit.edu
\and Ali Mili\thanks{This research is partially support by NSF Grant 
number DGE2043104}\\
New Jersey Institute of Technology \\
Newark, NJ USA\\
mili@njit.edu
} 

\maketitle

\runninghead{Wided Ghardallou et al.}{Invariant Relations}

\begin{abstract}
Great advances in program analysis would be enabled if it were possible to
derive the function of a program from inputs to outputs (or from initial
states to final states, depending on how we model
program semantics).  Efforts to do so have always stalled against the
difficulty to derive the function of loops; the expedient solution to
capture the function of loops by unrolling them an arbitrary number of
iterations is clearly inadequate.  In this paper, we propose a relations-based
method to derive the function of a C-like program, 
including programs that have loops nested
to an arbitrary level.  To capture the semantics of loops, we use the
concept of {\em invariant relation}.
\end{abstract}

\begin{keywords}
Invariant relations, Symbolic execution, 
while loops, Mathematica (\copyright
Wolfran Research),
Assume(), Capture(), Verify(), Establish().
\end{keywords}

\section{Introduction:  Analyzing C-like Programs}

Despite decades of research in programming language design and
implementation, and despite the emergence of several programming
languages that have advanced, sophisticated, technical attributes,
most software being developed today is written in C-like languages.
The six top languages in the September 2023 Tiobe 
Index\footnote{{\tt http://www.tiobe.com/}}
(C, C++, C\#, Java, Javscript, Python) are all derived from
or inspired by C.  The Tiobe Index reflects the frequency of
occurrence of each programming language in web searches, and is
usually interpreted as a measure of the frequency of use of each
language.  The dominance of C-like languages applies not only to
new
software development, but perhaps even more to the maintenance and
evolution of legacy software.

As software maintenance and evolution continue to account for a large,
and growing,
percentage of software engineering costs and resources, and as
software is increasingly developed from existing code, the 
ability to derive the function of a software artifact from
an analysis of its source code becomes increasingly critical.
The recent talk of using artificial intelligence to generate 
code makes this capability even more critical because AI generated
code bridges two gaps that ought to be validated
meticulously: the gap between
what the user intends and 
what the user formulates as a prompt; and
the gap (chasm?) between the user prompt and what the AI generator
understands it to mean / specify
\cite{greengard2023,sarkar2022,chenetal2022,liyi2021,feng2018,%
vasconcelos2022,welsh2023,yellin2023}.  This latter gap is all the more
critical that AI systems are notoriously opaque, thereby precluding
any process-based quality controls.

To derive the function of a C-like program, we use a standard 
parser to transform the program into an abstract syntax tree ({\em AST},
for short), then we perform a recursive descent on the tree to
derive an equation between the initial states and the final
states of the program;  this equation is formulated in the
syntax of {\em Mathematica} (\copyright Wolfran Research), and
captures the semantics of the program.  The function of the program
can be derived by solving this equation in the final state as a
function of the initial state.
The question of deriving the function of a 
program written in a C-like language has eluded researchers for decades,
primarily due to the presence of loops, whose function cannot be
easily modeled in general.  In this paper we show how we can,
under some conditions, capture the
function of iterative statements, such as while loops, for loops,
repeat loops, etc, at arbitrary levels of nesting.  
We do so by means of the concept of {\em invariant relation}, which is
a relation that links states that are separated by an arbitrary
number of iterations.

But deriving the function of a program in all its minute detail
may be too much information for an analyst to handle; programs
are complex artifacts, whose function may contain too much 
detail, not all of it is relevant to an analyst (for example,
how execution affects auxiliary variables does not matter as
much as the impact of the auxiliary variables on important state
variables).
Also, a programmer who abhors
poring over pages of source code will probably not relish the
prospect of poring over pages of mathematical notation instead.
Hence in addition to the ability to compute the function of a program,
we are interested to 
offer the user the ability 
to query parts of the program at arbitrary scale.  
To this effect, we propose four functions, which can be applied to
the source code of a program to analyze its semantic properties.
These functions (Assume(), Verify(), Capture(), Establish()) can 
refer to a program label or program part, and are used to make
assumptions or verify properties of program states or program
functions.

In section \ref{mathsect} we briefly introduce some definitions
and notations pertaining to relational mathematics; we assume that
the reader is familiar with these notions, our only purpose is to
specify notations and conventions.  
In section \ref{invsect} we introduce invariant relations,
discuss their properties, explore their use in the analysis of loops,
then discuss how we generate
them.  In section \ref{mapsect} we
discuss how we map programs into equations, and how we can
use these equations to analyze the semantics of a program.
In section \ref{querysect} we define four functions that can be
used to query the source code of a C-like program and 
illustrate their operation, then 
in section \ref{acvesect} we discuss their implementation 
in an interactive GUI (Graphic User Interface).  In section
\ref{concsect} we summarize our findings, compare them to
related work, discuss
threats to their validity, then sketch directions for future 
research.

\section{Mathematical Background}
\label{mathsect}

We assume the reader familiar with relational mathematics; the purpose of
this section is merely to introduce some definitions and
notations, inspired from \cite{brink1997}.

\subsection{Definitions and Notations}

We consider a set $S$ defined by the values of some program
variables, say $x$ and $y$; we denote elements
of $S$ by $s$, and we note that $s$ has the form
$s=\langle x, y\rangle.$
We denote the
$x$-component and (resp.) $y$-component of $s$
by $x(s)$ and $y(s)$.
For elements $s$ and $s'$ of $S$,
we may use $x$ to refer to $x(s)$ and $x'$
to refer to $x(s')$.  We refer to $S$ as the {\em space} of the
program and to $s \in S$ as a {\em state} of the program.
A relation on $S$ is a subset of the cartesian product
$S\times S$.
Constant relations on some set $S$
include
the {\em universal} relation,
denoted by $L$, the {\em identity} relation, denoted by $I$, and the
{\em empty} relation, denoted by $\emptyrelation$.

\subsection{Operations on Relations}
\label{operations}

Because relations are sets, we apply set theoretic operations
to them:  union ($\cup$), intersection
($\cap$), difference ($\setminus$), and complement ($\overline{R}$).
Operations on relations also include:
The {\em converse}, denoted by
$\widehat{R}$,
and
defined by $\widehat{R}=\{(s,s')| (s',s)\in R\}.$
The {\em product} of relations $R$ and $R'$ is the relation
denoted by $R\circ R'$ (or $RR'$)  and defined by $R\circ R'=\{(s,s')|
\exists s'': (s,s'')\in R\wedge (s'',s')\in R'\}.$
The $n^{th}$ {\em power} of relation $R$, for natural number $n$, is
denoted by $R^n$ and defined by $R^0=I$, and $R^n=R\circ R^{n-1}$, for
$n\geq 1$.
The {\em transitive closure} of relation $R$ is the relation denoted
by $R^+$ and defined by $R^+=\{(s,s')| \exists i>0: (s,s')\in R^i\}$.
The {\em reflexive transitive closure} of relation $R$ 
is the relation denoted by
$R^*$ and defined by $R^*=I\cup R^+$.  
Given a predicate $t$, we denote
by $T$ the relation
defined as $T=\{(s,s')| t(s)\}$.
The {\em pre-restriction} (resp. {\em post-restriction}) of relation
$R$
to predicate $t$ is the relation $T\cap R$ 
(resp. $R\cap\widehat{T}$). 
The {\em domain} of relation $R$ is defined as \(\dom(R)=\{s| \exists
s': (s,s')\in R\},\) and the {\em range} of $R$ ($\rng(R)$) is the domain of
$\widehat{R}$.
We apply the usual conventions for operator
precedence: unary operators are applied first, followed by
product, then intersection, then union.

\subsection{Properties of Relations.}

We say that $R$ is {\em deterministic} (or that it is a {\em
function})
if and only if $\widehat{R}R\subseteq I$,
and we say that $R$ is {\em total} if and only if $I\subseteq
R\widehat{R}$, or equivalently, $RL=L$; also, we say that
$R$ is {\em surjective} if and only if $LR=L$.
A {\em vector} $V$ is a relation that satisfies $VL=V$;
in set theoretic terms, a vector on
set $S$ has
the form $C\times S$, 
for some subset $C$ of $S$; we use vectors as relational
representations of sets.
We note that for a relation $R$, $RL$ represents the vector
$\{(s,s')| s\in\dom(R)\}$; we use $RL$ as the relational
representation of the domain of $R$.
A relation $R$ is said to be {\em reflexive}
if and only if $I\subseteq R$, {\em transitive} if and only
if $RR\subseteq R$ and {\em symmetric} if and only if
$R=\widehat{R}$.  
We admit without proof that the transitive closure of a relation $R$
is the smallest transitive superset of $R$ and that the reflexive
transitive closure of $R$ is the smallest reflexive transitive
superset of $R$.
A relation that is reflexive, symmetric and transitive is called
an {\em equivalence relation}.  
The following Proposition will be referenced throughout this paper.
\begin{proposition}
\label{eqfunprop}
Let $f$ and $g$ be two functions on space $S$.  Then $f=g$ if and only
if $f\subseteq g$ and $\dom(g)\subseteq\dom(f)$.
\end{proposition}

\begproof
Necessity is trivial.  For sufficiency, let $(s,s')$ be an
element of $g$; then $s$ is an element of $\dom(g)$, whence
$s$ is an element of $\dom(f)$.  From $(s,f(s))\in f$, we infer
that $(s,f(s))\in g$; since $g$ is deterministic, $f(s)=s'$.
Hence $(s,s')\in f$.  From $g\subseteq f$ and $f\subseteq g$
we infer $f=g$.
\finproof

\subsection{Loop Semantics}

Given a program {\tt p} on state space $S$, the {\em function}
of program {\tt p} is denoted by $P$ and defined as the set of
pairs of states $(s,s')$ such that if {\tt p} starts execution
in state $s$ it terminates normally (i.e. after a finite number
of steps, without attempting any illegal operation) in state $s'$.
As a result, the domain of $P$ is the 
set of states $s$ such that if {\tt p} is executed on state
$s$ it terminates normally.  
By abuse of notation, we may sometimes
denote a program {\tt p} and its function $P$ by the same symbol, $P$.

The following Proposition gives a formula for the function of a loop.
\begin{proposition}
\label{loopfunctionprop}
The function of the while loop {\small{\tt w: \{while (t) \{b\}\}}} on space
$S$ is:  
$$W=(T\cap B)^*\cap\widehat{\overline{T}},$$ 
where $T$ is the
vector that represents condition {\tt t} and $B$ is the function of 
{\tt \{b\}}.
\end{proposition}
A proof of this Proposition is given in \cite{mili2012scp}.

\section{Invariant Relations}
\label{invsect}

In this section, we briefly present a definition of invariant relations,
then we discuss in turn:  what we use invariant relations for; 
then how we generate invariant relations. 

\begin{definition}
\label{invreldef}
Given a while loop $w$ of the form {\small{\tt \{while (t) \{b\}\}}},
an {\em invariant relation} of $w$ is a reflexive transitive superset
of $(T\cap B)$, where $T$ is the vector that represents condition
{\small{\tt t}} and $B$ is the function of {\small{\tt\{b\}}}.
\end{definition}
The following properties stem readily from the definition of
invariant relations.
\begin{proposition}
Given a while loop {\tt w: \{while (t) \{b;\}\}} on space $S$,
\begin{itemize}
\item The intersection of invariant relations of {\tt w}
is an invariant relation of {\tt w}.
\item $(T\cap B)^*$ is an invariant relation of {\tt w}.
\item $(T\cap B)^*$ is the smallest invariant relation of {\tt w}:
it is a subset of any invariant relation of {\tt w}.
\end{itemize}
\end{proposition}

\begproof
The first clause is a consequence of the property that the
intersection of reflexive transitive relations is reflexive and
transitive.  The second and third clauses are properties of
the reflexive transitive closure of a relation.
\finproof

For illustration, we consider space $S$ defined by natural variables
$n$, $f$, $k$, and we let {\tt w} be the following program:{\small
\begin{verbatim}
   w:  {while (k!=n+1) {f=f*k; k=k+1;}}
\end{verbatim}}

\noindent
For this program we find (using $s$ as a stand-in for 
$\langle n,f,k\rangle$ and $s'$ as a stand-in for
$\langle n',f',k'\rangle$):\\
\tabeq $T=\{(s,s')| k\neq n+1\}$,\\
\tabeq $B=\{(s,s')| n'=n\wedge f'=f\times k\wedge k'=k+1\}$.\\
We leave it to the reader to check
that the following are invariant relations for $w$:\\
\tabeq $R_0=\{(s,s')|n'=n\}$,\\
\tabeq $R_1=\{(s,s')|k\leq k'\}$,\\
\tabeq $R_2=\{(s,s')| \frac{f}{(k-1)!}=\frac{f'}{(k'-1)!}\}$.\\
Since
these are invariant relations of $w$, so is their intersection $R=
R_0\cap R_1\cap R_2$.

\subsection{Using Invariant Relations}

The following Proposition is a
simple corollary of Proposition \ref{loopfunctionprop};
it stems readily from the property that any invariant relation
$R$ of {\tt w} is a superset of $(T\cap B)^*$.
\begin{proposition}
\label{wrtprop}
Given a while loop of the form {\small{\tt w: \{while (t) \{b\}\}}} 
on space
$S$, and an invariant relation $R$ of {\tt w} the following inequality holds:
$W\subseteq R\cap\widehat{\overline{T}}$.
\end{proposition}
As an illustration,
we consider the factorial program above and we compute the
proposed upper bound $W'$ of $W$:\\
\tabeq $W'=R\cap\widehat{\overline{T}}$\\
= \tabcom \{substitution, simplification\}\\
\tabeq $\{(s,s')| k\leq n+1\wedge n'=n\wedge k'=n+1
\wedge f'=n!\times \frac{f}{(k-1)!}\}$.\\
According to Proposition \ref{wrtprop}, $W$ is a subset
of $W'$ (given above).  On the other hand, the loop
converges for all $s$ that satisfy the condition $(k\leq n+1)$,
hence $\dom(W')\subseteq \dom(W)$.  Since $W'$ is deterministic,
we can apply Proposition
\ref{eqfunprop}, which yields $W=W'$.
This is indeed the function of the (uninitialized) loop.


\subsection{The Elementary Invariant Relation}

The following Proposition gives
a constructive / explicit 
formula for a trivial invariant relation, which
we call the {\em elementary invariant relation}.

\begin{proposition}
\label{elemprop}
Due to \cite{jsc2012}.
Given a while loop of the form {\small{\tt w: 
\{while (t) \{b\}\}}} on space
$S$, the following relation is an invariant relation for {\tt w}:
$$E=I\cup T(T\cap B),$$ 
where $T$ is the vector that represents the
loop condition and $B$ is the function of the loop body.
\end{proposition}

This relation includes pairs of states
$(s,s')$ such that $s=s'$, reflecting the case when the loop does
not iterate at all, and pairs of states $(s,s')$ when the loop iterates
at least once, in which case $s$ satisfies $t$ and $s'$ is in the
range of $(T\cap B)$.

As an illustration, we consider the factorial loop introduced above,
in which we change the condition from {\tt t:(k!=n+1)} to
{\tt t':(k<n+1)}:{\small
\begin{verbatim}
   v:  {while (k<n+1) {f=f*k; k=k+1;}}
\end{verbatim}}
This change does not affect the loop's
invariant relations.  If we apply the formula of Proposition \ref{wrtprop}
to the invariant relation $R=R_0\cap R_1\cap R_2$, 
we find the following upper bound of $V$:\\
\tabeq $R\cap\widehat{\overline{T'}}$\\
=\tabcom\{Substitutions\}\\
\tabeq $\{(s,s')| k\leq k'\wedge n'=n\wedge 
\frac{f}{(k-1)!}=\frac{f'}{(k'-1)!}\wedge k'\geq n+1\}$.\\
This is not a deterministic relation:  unlike in the
previous example (where we had $k'=n+1$) we do not have sufficient
information to derive the final value of $k'$.
The elementary invariant relation will help us derive the
missing information:\\
\tabeq $E=I\cup T'(T'\cap B)$\\
=\tabcom\{Substitutions\}\\
\tabeq $I\cup \{(s,s')| k<n+1\wedge \exists s'':
k''<n''+1\wedge n'=n''\wedge k'=k''+1\wedge f'=f''\times k''\}$\\
=\tabcom\{Simplification\}\\
\tabeq $I\cup \{(s,s')| k<n+1\wedge k'<n'+2\wedge \exists f'': 
f'=f''\times(k'-1)\}$.\\
Applying Proposition \ref{wrtprop} with the new invariant relation
$R'=R\cap E$ yields the following upper bound $V'$ for $V$,
where we abbreviate $(\exists f'': f'=f''\times(k-1))$ by
$\kw{mult}(f',k'-1)$:\\
\tabeq $V'=R_0\cap R_1\cap R_2\cap E\cap\widehat{\overline{T'}}$\\
=\tabcom\{Factoring $E$\}\\
\tabeq $I\cap R_0\cap R_1\cap R_2\cap\widehat{\overline{T'}}$\\
\tabeq $\cup$\\
\tabeq $R_0\cap R_1\cap R_2\cap\{(s,s')|k<n+1\wedge k'<n+2\wedge
\kw{mult}(f',k'-1)\}
\cap\widehat{\overline{T'}}$\\
=\tabcom\{Substitutions, Simplifications\}\\
\tabeq $I\cap \overline{T'}$\\
\tabeq $\cup\{(s,s')|k<n+1\wedge n'=n\wedge
k'\geq n+1\wedge \frac{f}{(k-1)!}=\frac{f'}{(k'-1)!}
\wedge k'<n+2\wedge
\kw{mult}(f',k'-1)\}$\\
=\tabcom\{Simplifications, making $\kw{mult}(f',k'-1)$ redundant\}\\
\tabeq $\{(s,s')| k\geq n+1\wedge n'=n\wedge k'=k\wedge f'=f\}$\\
\tabeq $\cup\{(s,s')|k<n+1\wedge n'=n\wedge
k'=n+1\wedge f'=n!\times\frac{f}{(k-1)!}\}$.\\
We see that $V'$ is deterministic; on the other hand, the loop
converges for all initial states, 
hence $\dom(V)=S$,
from which we infer vacuously:
$\dom(V')\subseteq \dom(V)$.  By Proposition
\ref{eqfunprop}, we infer $V'=V$; the elementary invariant
relation was instrumental in enabling
us to compute the function of the loop.

\subsection{Generation of Invariant Relations}

The elementary invariant relation is the only invariant
relation
we get for free, so to speak:  we derive it constructively
from the features of the loop, $T$ and $B$; not surprisingly,
it gives very little (yet sometimes crucial) information about
the function of the loop.  All the other invariant relations
can only be derived by a meticulous analysis of $T$ and $B$.
The following Proposition is the basis of our approach to the
generation of invariant relations.

\begin{proposition}
Given a while loop of the form {\small{\tt w:\{while (t) \{b\}\}}} 
on space
$S$, and a superset $B'$ of $(T\cap B)$, then $B'^*$ is an invariant relation
for $w$.
\end{proposition}

\begproof
If $B'$ is a superset of $(T\cap B)$, then so is $B'^*$; on the other
hand, $B'^*$ is by construction reflexive and transitive.
\finproof

To generate invariant relations, it suffices to isolate supersets
of the function of the guarded loop body, and generate reflexive
transitive supersets thereof.  In
particular, 
if we decompose $(T\cap B)$ as the intersection of several terms,
say:
$$(T\cap B)=B_1\cap B_2\cap B_3 ... \cap B_n,$$
then we can compute the reflexive transitive closure 
(say $R_i$) of each $B_i$ and take their
intersection to find an invariant relation for the loop as:
$R=\bigcap_{i=1}^n R_i.$

\subsection{A Database of Recognizers}

In order to automate the process of invariant relation generation,
we create a database of {\em recognizers}, where each recognizer is
made up of four components:
\begin{itemize}
\item {\em Formal Space Declaration, $\sigma$}.  This includes C-like variable
declarations and constant declarations (as needed).
These are intended to be matched against actual variable declarations
of the loop.
\item {\em Condition of Application, $\alpha$}.  
Some recognizers can be applied
only under some conditions, which we formulate in this column.
\item {\em Formal Clause, $\gamma$}.  This includes clauses that are
formulated in terms of the variables declared above, and
are intended to be matched against the function of the
guarded loop body.
\item {\em Invariant Relation Template, $\rho$}.  This represents 
a superset of the
reflexive transitive closure of the formal clause $\gamma$ (above).
\end{itemize}
The following pattern-matching algorithm is deployed to generate
invariant relations of a loop, using the database of recognizers:
\begin{itemize}
\item For each recognizer in the database,
\begin{itemize}
\item For each type-compatible mapping $m$ from the formal variable
declarations of the recognizer to the actual variable declarations 
of the program,
\begin{itemize}
\item We check whether condition $m(\alpha)$ holds (i.e. the recognizer
is applicable) and whether $F$ logically implies $m(\gamma)$
(i.e. the function of the guarded loop body matches the formal
clause of the recognizer, in which formal variable names are replaced
by actual variable names).  
\item In case of success, we generate $m(\rho)$; in other words,
we instantiate the invariant relation template (formulated in 
formal variable names) with actual variable names.  This gives
us an invariant relation of the loop.
\end{itemize}
\end{itemize}
\end{itemize}
Table \ref{rectab} shows some sample recognizers, for purposes 
of illustration,
where $\concat$ represents list concatenation and also
(by abuse of notation) the operation of appending an
element to a list; and $Fib()$ represents the
Fibonacci function.  We distinguish between 1-recognizers, 2-recognizers and
3-recognizers, depending on the number of conjuncts in their
{\em Formal Clause} component ($\gamma$); in practice we find
that we seldom need to match more than three conjuncts at a time.

\begin{table}
\begin{center}
{\footnotesize
\begin{tabular}{|l|l|l|l|l|}
\hline\hline
ID & \shortstack[l]{Formal\\ Space, $\sigma$} & 
\shortstack[l]{Condition,\\ $\alpha$} &
\shortstack[l]{Formal \\Clause, $\gamma$} & 
\shortstack[l]{Invariant Relation\\ Template, $\rho$}\\
\hline\hline
1R1 & {\tt AnyType x;} & \truekw & $x'=x$ & $x'=x$\\
\hline
1R2 & {\tt int i;} & \truekw & $i'=i+1$ & $i\leq i'$ \\
\hline
1R3 & \shortstack[l]{{\tt int x;}\\
{\tt const int a}}& $a\neq 0$ & $x'=x+a$ &\shortstack[l]{$ax\leq ax'\wedge$\\
$x \modkw |a|=x' \modkw |a|$}\\
\hline\hline
2R0 & \shortstack[l]{{\tt int x, y;}\\
{\tt const int a, b}}& true & \shortstack{$x'=x+a$\\
$\wedge y'=y+b$}& $ay-bx=ay'-bx'$\\
\hline
2R2 & {\tt listType v,w} &
$length(v)>0$ &
\shortstack[l]{$v'=\tail(v)\wedge$\\
$w'=w\concat\head(v)$} &
\shortstack[l]{$v'\leq v\wedge$\\
$w\concat v=w'\concat v'$}\\
\hline
2R3 & \shortstack[l]{{\tt listType v;}\\{\tt int x;}\\
{\tt const int a}} & $a\neq 0$ & 
\shortstack[l]{$v'=tail(v)\wedge$\\$x'=x+a$} &
\shortstack[l]{$x+a\times\length(v)$\\
\tabeq$=x'+a\times\length(v')$}\\
\hline
2R4 & \shortstack[l]{{\tt int f(int i);}\\{\tt int i,x;}} &
$i\geq 0$ & \shortstack[l]{$i'=i+1\wedge$\\$x'=x+f(i)$} &
\shortstack[l]{$x-\Sigma_{k=0}^{i-1}f(k)$\\
\tabeq$=x'-\Sigma_{k=0}^{i'-1}f(k)$}\\
\hline\hline
3R1 & 
\shortstack[l]{{\tt const int N;}\\{\tt real x, a[N];}\\
{\tt int i}} & 
$N\geq 0$ &
\shortstack[l]{$x'=x+a[i]$\\$\wedge i'=i-1$\\$\wedge a'=a$} &
\shortstack[l]{$x+\Sigma_{k=0}^{i}a[k]$\\
\tabeq $=x'+\Sigma_{k=0}^{i'}a'[k]$}\\
\hline
3R2 & {\tt int x, y, i} & true &
\shortstack[l]{$x'=x+y\wedge$\\$y'=x\wedge$\\$i'=i+1$} &
\shortstack[l]{
$x'=x\times Fib(i'-i+1)$\\$\tabeq +y\times Fib(i'-i)$\\
$y'=x\times Fib(i'-i)$\\$\tabeq +y\times Fib(|i'-i-1|)$}\\
\hline\hline
\end{tabular}}
\caption{\label{rectab}Sample Recognizers}
\end{center}
\end{table}

\subsection{Condition of Sufficiency}

Given a while loop {\tt w:\{while (t) \{b;\}\}} that we wish to
analyze, we compute its invariant relation $E=I\cup T(T\cap B)$,
then we use recognizers such as those of Table \ref{rectab} to
generate further invariant relations, take their intersection with $E$,
say $R$, then use Proposition \ref{wrtprop} to derive a superset
(approximation) $W'$ of the function $W$ of {\tt w}.  As we find
more and more invariant relations and take their intersection, 
the resulting invariant relations grows smaller and smaller.  The question
we ask is:  how small does $R$ have to be to ensure that we actually
find the exact function $W$.  It is tempting to think that in
order to derive the function of the loop ($W=(T\cap B)^*\cap
\widehat{\overline{T}}$) by means of the approximation given 
in Proposition \ref{wrtprop} ($W'=R\cap\widehat{\overline{T}}$),
it is necessary to derive the smallest invariant relation
($R=(T\cap B)^*$).  According to the following Proposition,
this is not necessary.
\begin{proposition}
\label{sufficiencyprop}
Given a while loop {\tt w:\{while (t) \{b;\}\}} on space $S$
and an invariant relation $R$ of {\tt w}, the function $W$ of {\tt w}
equals $R\cap\widehat{\overline{T}}$ if and only if:
\begin{itemize}
\item $R\cap\widehat{\overline{T}}$ is deterministic.
\item $\dom(R\cap\widehat{\overline{T}})\subseteq\dom(W)$.
\end{itemize}
\end{proposition}

\begproof
Necessity is trivial.  For sufficiency, consider that because of the
first hypothesis, we can use Proposition \ref{eqfunprop}, since
$R\cap\widehat{\overline{T}}$ and $W$ are both deterministic.
The condition $W\subseteq R\cap\widehat{\overline{T}}$ stems from
Proposition \ref{wrtprop}; the condition 
$\dom(R\cap\widehat{\overline{T}})\subseteq\dom(W)$ is given by 
hypothesis.  By Proposition \ref{eqfunprop}, $W=R\cap\widehat{\overline{T}}$.
\finproof

Henceforth we use $W$ to designate the function of the loop and
$W'$ to designate the approximation $W'=R\cap\widehat{\overline{T}}$
that we can derive from some
invariant relation $R$; we rename $W'$ as $W$ once we ensure that
$W'$ does satisfy the conditions of 
Proposition \ref{sufficiencyprop}.

\begin{figure}
\thicklines
\setlength{\unitlength}{0.015in}
\begin{center}
\begin{picture}(200,100)

\put(40,0){\framebox(160,60)}
\put(20,20){\framebox(140,60)}
\put(0,20){\framebox(160,80)}
{\scriptsize
\put(180,10){\makebox(0,0){$\widehat{\overline{T}}$}}
\put(100,40){\makebox(0,0){$(T\cap B)^*\cap\widehat{\overline{T}}=
R\cap\widehat{\overline{T}}$}}
\put(40,70){\makebox(0,0){$(T\cap B)^*$}}
\put(10,90){\makebox(0,0){$R$}}
}
\end{picture}
\end{center}
\caption{\label{starkfig}Achieving $(T\cap B)^*\cap\widehat{\overline{T}}
=R\cap\widehat{\overline{T}}$ does not require $R=(T\cap B)^*$}
\end{figure}
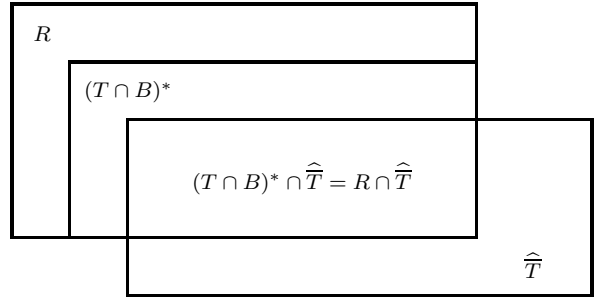

Given that we are approximating $W=(T\cap B)^*\cap\widehat{\overline{T}}$
by $W'=R\cap\widehat{\overline{T}}$
for some invariant relation $R$ of {\tt w}, it is tempting to think that we can
achieve the condition $W=W'$ only with 
the invariant relation $R=(T\cap B)^*$.  
But this Proposition
makes no mention of such a condition,
and the following (counter-)
example, due to \cite{desharnais2023}, shows that
we can achieve $W=R\cap\widehat{\overline{T}}$ while $R$ is not
equal to $(T\cap B)^*$, but is a superset thereof.
\begin{itemize}
\item {\em Space}, $S=\{0,1,2\}$.
\item {\em Loop Condition}, 
$T=\{1,2\}\times S$.  Whence, $\widehat{\overline{T}}=S\times \{0\}$.
\item {\em Loop Body Function}, $B=\{(1,0),(2,0)\}$.
\item {\em Invariant Relation},
$R=\{(0,0),(1,0),(1,1),(1,2),(2,0),(2,1),(2,2)\}$.
\end{itemize}
We can easily check:\\
\tabeq $T\cap B=\{(1,0),(2,0)\}$; \\
whence\\
\tabeq 
$(T\cap B)^+=(T\cap B)$,\\
from which we infer\\
\tabeq $(T\cap B)^*=I\cup(T\cap B)=\{(0,0),(1,0),(1,1),(2,0),(2,2)\}$.\\
Taking the intersection with $\widehat{\overline{T}}$, we find:\\
\tabeq $W=(T\cap B)^*\cap\widehat{\overline{T}}=\{(0,0),(1,0),(2,0)\}$.\\
Now we consider the approximation $W'=R\cap\widehat{\overline{T}}$ derived
from relation $R$; first, we must check that $R$ is indeed an invariant
relation.  Clearly, it is reflexive; on the other hand, we verify
easily that $R^2\subseteq R$, hence $R$ is transitive;  finally it is also a
superset of $(T\cap B)$, hence $R$ is an invariant relation.  We
compute:\\
\tabeq $W'=R\cap\widehat{\overline{T}}$\\
=\tabcom\{Substitutions\}\\
\tabeq $\{(0,0),(1,0),(1,1),(1,2),(2,0),(2,1),(2,2)\}\cap
\{(0,0),(1,0),(2,0)\}$\\
=\tabcom\{Inspection\}\\
\tabeq $(T\cap B)^*\cap\widehat{\overline{T}}$.\\
Hence we have shown on this example that the function of the loop
$W=(T\cap B)^*\cap\widehat{\overline{T}}$ can be equal to the approximation
derived from an invariant relation $R$, namely
$W'=R\cap\widehat{\overline{T}}$ even though $R$ is not equal to
$(T\cap B)^*$, but is a superset thereof:\\
\tabeq $(T\cap B)^*=\{(0,0),(1,0),(1,1),(2,0),(2,2)\}$,\\
\tabeq $R=\{(0,0),(1,0),(1,1),(1,2),(2,0),(2,1),(2,2)\}$.\\
Figure \ref{starkfig} shows in stark, plain terms why this is
hardly surprising: just because two sets have the same intersection
with a third set does not mean they are equal.  

{\bf{\em The practical implication of this result is that we can compute
the function of a while loop without having to compute the reflexive
transitive closure of its guarded loop body}}.  In other words, we
do not need to find the smallest invariant relation of a while loop;
rather it suffices to find a small enough invariant relation $R$
to satisfy the conditions of Proposition
\ref{sufficiencyprop} (both of which impose upper bounds on $R$).
This suggests that authors who approximate the function of a loop
by computing the transitive closure of its loop body function may
be aiming unnecessarily high, i.e. setting a goal that is unnecessarily
difficult to achieve.

\section{Mapping Programs to Program Functions}
\label{mapsect}
\label{semanticssect}
\label{threestepsect}


\subsection{A Three Step Process}

In this section we discuss how we derive
the function of a program, written in the
syntax of Mathematica, from its Java source code.  Our tool
proceeds in three steps:
\begin{itemize}
\item {\bf J2A}:  {\em From Java to an Abstract Syntax Tree}.  
In this step, we deploy a standard Java parser that analyzes
the source code for syntactic correctness and maps it onto an
abstract symtax tree.  
All subsequent steps are independent of the
source language; so that we can handle any C-like language
for which we can find a parser that maps source code onto
an AST of the right format.
\item {\bf A2M}:  {\em From the AST to Mathematica} (\copyright 
Wolfram Research).  Each internal node of the Abstract Syntax Tree
represents a control structure of the programming language, and each
leaf represents an elementary programming language construct (such
as an assignment statement, a variable declaration, etc).
Our goal is to map the abstract syntax tree into an equation between
the initial state of the program (represented by, say, state $s$)
and its final state (represented by $sP$).  We do so by a recursive
descent through the abstract syntax tree, where at each node we
generate a Mathematica equation that captures the semantics of
the programming construct represented by that node.
\item {\bf M2F}: {\em From a Mathematica Equation to a Program Function}.
Regardless of how large or how deeply nested, the Mathematica
equation generated in the previous step is an equation between
the initial state and the final state of the program; in the
case of the sample program of Figure \ref{codefig}, it is an
equation involving variables $\langle x,y,t,i,j,k\rangle$, which
represent the initial state and variables 
$\langle xP,yP,tP,iP,jP,kP\rangle$, which represent the final
state of the program.  In this step, we invoke Mathematica to
derive the final state as a function of the initial state,
thereby yielding the function of the program.  In other words, we
transition from an equation of the form $Eq(s,sP)$ to a 
solution of the form $sP=F(s)$; $F$ is then the function of the
program. 
\end{itemize}
The first step (J2A) is carried out by a routine Java parser; the
third step (M2F) is carried out by Mathematica.
Our contribution is in step A2M, wich we discuss in the next
subsection.

\subsection{A Recursive Descent}

The abstract syntax tree includes two types of leaves, i.e. nodes
which do not give rise to any recursive calls:  variable declarations
and assignment statements.
\begin{itemize}
\item {\em Variable Declaration}.  The effect of a variable declaration
is to add the new variable in the symbol table while preserving the
values of all the current variables.  Hence we write:\\
\tabeq {\tt A2M(var x) = (x1==x1P \&\& x2==x2P \&\& .. xn==xnP),}\\
for all current state variables {\tt x1... xn}.
The symbol table includes the name of the variable and its data type,
as well as the primed version of the variable, and the double primed
version, as shown below:\\
\begin{center}
\begin{tabular}{|l|l|l|l|}
\hline\hline
var & type & primed & 2-primed\\
\hline\hline
x & int & xP & xPP\\
\hline
... & ... & ... & ...\\
\hline\hline
\end{tabular}
\end{center}
One version we are considering would assign an undefined value to the
newly declared variable, and we should then extend the semantic definition
of expressions
in such a way that when it appears in any expression, this undefined value
causes the expression to take the undefined value; this would be useful
to alert users to unassigned variables
(i.e. cases where a variable is referenced before it is given
a value).  Such a rule would then be 
written as:\\
\tabeq {\tt A2M(var x) = (xP==Undef \&\& 
x1P==x1 \&\& x2P==x2 \&\& .. xnP==xn),}\\
\item {\em Assignment Statement}.  Given a variable {\tt x} and an
expression {\tt E} whose value is compatible with the 
data type of {\tt x},
we let {\tt defE(s)} be the predicate that specifies the condition under
which expression $E$ can be evaluated at state {\tt s}, then:\\
\tabeq {\tt A2M(x=E) = (defE(s) \&\& xP==E(s) \&\& x1P==x1
\&\& x2P==x2 \&\& .. xnP==xn).}\\
\end{itemize}

The internal nodes of the Abstract Syntax Tree give rise to
recursive calls of {\tt A2M()}; we specify below the result
of applying {\tt A2M()} to each type of internal node.
\begin{itemize}
\item {\em Block}.  We consider a simple scope, consisting of a single
variable declaration within brackets.  We write:\\
\tabeq {\tt A2M(\{Xtype x; B\})=[Exists x, xP: A2M(B)]}.\\
Let $S$ be the space of the program outside the block; then the
space of {\tt B} is the cartesian product of $Xtype$ by $S$ whereas
the space of {\tt \{Xtype x; B\}} is merely $S$.  Hence the function
of {\tt\{Xtype x; B\}} is the projection of the function of {\tt B} on
space $S$; this is achieved by quantifying ($\exists$) variables
{\tt x} and {\tt xP}.
\item {\em Sequence}.  We consider a sequence of two statements:\\
\tabeq {\tt A2M(p1;p2) = [Exists sPP: sP2sPP(A2M(p1)) \&\&
s2sPP(A2M(p2))],}\\
where {\tt sp2sPP(Eq)} replaces each instance of a primed variable
in Eq by a double primed variable, and {\tt s2spp(Eq)}
replaces each instance of an unprimed variable in Eq by a double
primed variable.  Consider that {\tt A2M(p1)} is an equation
in $s$ and $sP$, and {\tt sp2spp(A2M(p1))} is an equation in
$s$ and $sPP$; likewise, {\tt s2sPP(A2M(p2))} is an equation
in $sPP$ and $sP$.  By taking the conjunction of these two
equations and quantifying for $sPP$, we find an equation in
$s$ and $sP$ that relates the states before ($s$) and after ($sP$)
execution of
{\tt \{p1;p2\}}.
\item {\em If-then-else}.  The following rule stems readily from 
the semantics of the statement:\\
\tabeq {\tt A2M(if (t) A else B) = t \&\& A2M(A) || !t \&\& A2M(B)}.
\item {\em If-then}.  The following rule stems readily from 
the semantics of the statement:\\
\tabeq {\tt A2M(if (t) A) = t \&\& A2M(A) || !t \&\& sP==s}.
\end{itemize}
In the next section, we discuss the semantic definition of 
some common iterative constructs.

\subsection{Semantic Rules for Iteration}

Given a while loop of the form {\tt w: while (t) \{b;\}}, we let
{\tt invR(t,B)}, where $B$ is the function of the loop body,
be the intersection of all the invariant relations
that are generated for {\tt w}; this includes the elementary
invariant relation as well as the invariant relations that are
generated by pattern matching against the recognizers.
Using {\tt invR()}, we can define the semantics of any iterative
construct in C-like languages (while loop, for loop, repeat until,
do while, etc).  We present below the semantics of some 
common iterative constructs:
\begin{itemize}
\item {\em While Loop}.  
The following equation is a direct consequence of Proposition
\ref{wrtprop}.\\
\tabeq {\tt A2M(while (t) \{B\}) = invR(t,A2M(B)) \&\& !t(sP)}.\\
We can have loops nested to an arbitrary degree; the call and return 
sequence of the recursive descent ensures that inner loops are replaced
by their function before outer loops are analyzed.
\item {\em For Loop}. We use a simple for-loop pattern, for illustration:\\
\tabeq {\tt A2M(for (int i=a;i<b;i++) \{B\})}\\
=\\
{\tt [Exists i, iP: i==a \&\& iP==b \&\& invR(i<b, iP=i+1\&\& 
A2M(B))]}.\\
Note that $i$ is not a state variable; it is an auxiliary variable
that serves to count the number of iterations and is discarded, by
virtue of being quantified ({\tt Exists i, iP}).
This rule specifies the initial and final value of variable $i$,
and invokes the invariant relation generator {\tt invR()} with the
loop condition {\tt (i<b)}, and with the function of the loop body
{\tt A2M(B)} accompanied, so to speak, with the incrementation of
$i$ {\tt (iP=i+1}).
\item {\em Repeat Loop}.  A repeat loop is similar to a while
loop (with negated condition), except that the loop body is executed
before the loop condition is tested.  Whence the following rule:\\
\tabeq {\tt A2M(repeat \{B\} until (t))=
A2M(B; while (!t) \{B\})}.
\end{itemize}

\subsection{An Illustrative Example}

\begin{figure*}
\thicklines
\setlength{\unitlength}{0.030in}
\begin{center}{\bf{\tt{\scriptsize
\begin{picture}(140,100)
\put(20,100){\makebox(0,0){\framebox{;}}}
\put(18,98){\line(-5,-2){15}}
\put(22,98){\line(5,-2){16}}
\put(0,90){\makebox(0,0){\framebox{int y, x, i , t, j ,k}}}
\put(20,80){\makebox(0,0){\framebox{t=i-j}}}
\put(40,90){\makebox(0,0){\framebox{;}}}
\put(38,88){\line(-5,-2){15}}
\put(42,88){\line(5,-2){15}}
\put(60,80){\makebox(0,0){\framebox{if-else}}}
\put(58,78){\line(-4,-1){26}}
\put(62,78){\line(4,-1){23}}
\put(60,78){\line(0,-1){6}}
\put(30,70){\makebox(0,0){\framebox{;}}}
\put(28,68){\line(-1,-1){6}}
\put(32,68){\line(1,-1){6}}
\put(60,70){\makebox(0,0){\framebox{i>j}}}
\put(90,70){\makebox(0,0){\framebox{if-else}}}
\put(88,68){\line(-2,-1){14}}
\put(90,68){\line(0,-1){6}}
\put(92,68){\line(2,-1){14}}
\put(20,60){\makebox(0,0){\framebox{x=0}}}
\put(40,60){\makebox(0,0){\framebox{;}}}
\put(38,58){\line(-1,-1){6}}
\put(42,58){\line(1,-1){6}}
\put(70,60){\makebox(0,0){\framebox{while}}}
\put(68,58){\line(-1,-1){6}}
\put(72,58){\line(1,-1){6}}
\put(90,60){\makebox(0,0){\framebox{i<j}}}
\put(110,60){\makebox(0,0){\framebox{else}}}
\put(108,58){\line(-1,-1){6}}
\put(112,58){\line(1,-1){6}}
\put(30,50){\makebox(0,0){\framebox{y=f(x)}}}
\put(50,50){\makebox(0,0){\framebox{while}}}
\put(48,48){\line(-1,-1){6}}
\put(52,48){\line(1,-1){6}}
\put(62,50){\makebox(0,0){\framebox{j!=i}}}
\put(80,50){\makebox(0,0){\framebox{;}}}
\put(78,48){\line(-1,-1){6}}
\put(82,48){\line(1,-1){6}}

\put(120,50){\makebox(0,0){\framebox{;}}}

\put(40,40){\makebox(0,0){\framebox{i!=j}}}
\put(60,40){\makebox(0,0){\framebox{;}}}
\put(58,38){\line(-1,-1){6}}
\put(62,38){\line(1,-1){6}}
\put(70,40){\makebox(0,0){\framebox{j=j+k}}}
\put(90,40){\makebox(0,0){\framebox{;}}}
\put(100,50){\makebox(0,0){\framebox{while}}}
\put(100,48){\line(-1,-1){6}}
\put(105,48){\line(6,-1){35}}

\put(50,30){\makebox(0,0){\framebox{i=i+k}}}
\put(70,30){\makebox(0,0){\framebox{;}}}
\put(68,28){\line(-1,-1){6}}
\put(72,28){\line(1,-1){6}}
\put(85,30){\makebox(0,0){\framebox{k=k+1}}}
\put(100,30){\makebox(0,0){\framebox{;}}}
\put(88,38){\line(-1,-1){6}}
\put(92,38){\line(1,-1){6}}
\put(94,20){\makebox(0,0){\framebox{j=j-k}}}
\put(108,20){\makebox(0,0){\framebox{;}}}
\put(98,28){\line(-1,-1){6}}
\put(102,28){\line(1,-1){6}}
\put(108,18){\line(-1,-1){6}}
\put(110,18){\line(1,-2){3}}
\put(100,10){\makebox(0,0){\framebox{y=f(y)}}}
\put(112,10){\makebox(0,0){\framebox{;}}}

\put(60,20){\makebox(0,0){\framebox{k=k-1}}}
\put(80,20){\makebox(0,0){\framebox{;}}}
\put(78,18){\line(-1,-1){6}}
\put(82,18){\line(1,-1){6}}
\put(70,10){\makebox(0,0){\framebox{i=i-k}}}
\put(90,10){\makebox(0,0){\framebox{;}}}
\put(88,8){\line(-1,-1){6}}
\put(92,8){\line(1,-1){6}}
\put(80,0){\makebox(0,0){\framebox{y=f(y)}}}
\put(100,0){\makebox(0,0){\framebox{;}}}

\put(98,40){\makebox(0,0){\framebox{t!=i}}}
\put(140,40){\makebox(0,0){\framebox{;}}}
\put(138,38){\line(-1,-1){6}}
\put(142,38){\line(4,-1){20}}
\put(132,30){\makebox(0,0){\framebox{for}}}
\put(130,28){\line(-1,-2){13}}
\put(135,28){\line(2,-1){10}}
\put(162,30){\makebox(0,0){\framebox{y=x-y}}}
\put(150,20){\makebox(0,0){\framebox{x=x+1}}}
\put(128,0){\makebox(0,0){\framebox{(int z=0;z!=y;z++)}}}
\put(164,28){\line(-1,-1){6}}
\put(168,28){\line(1,-1){6}}
\put(162,20){\makebox(0,0){\framebox{t=t+1}}}
\put(174,20){\makebox(0,0){\framebox{;}}}
\end{picture}}}}
\caption{\label{astfig}AST:  Abstract Syntax Tree}
\end{center}
\end{figure*}
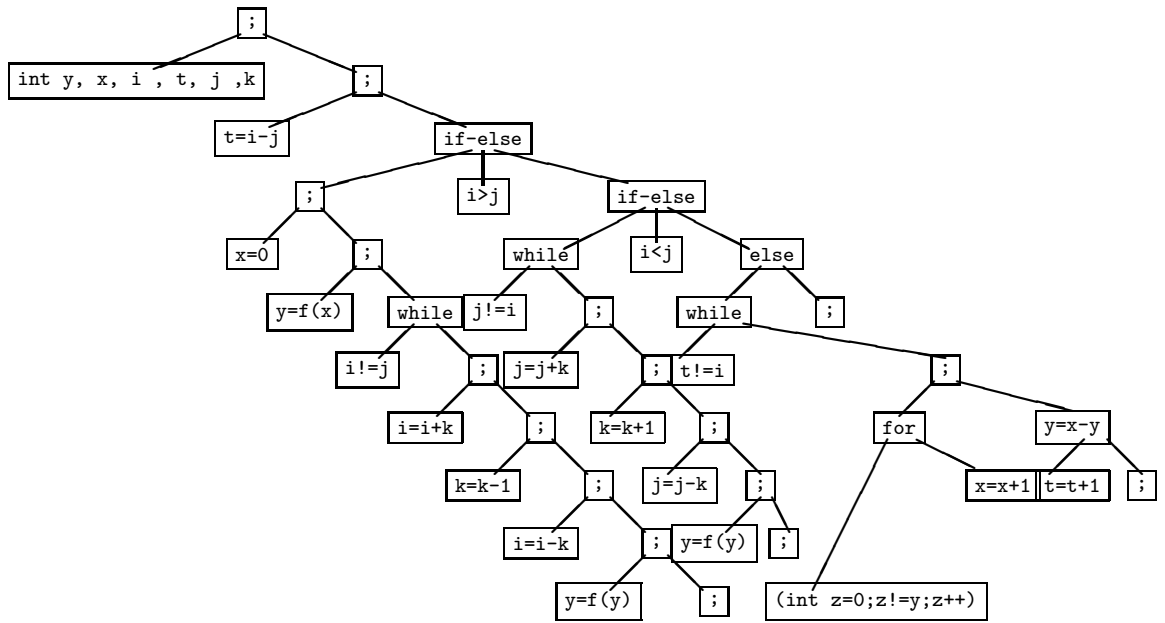

For the sake of illustration, we show in this section a sample
program written in Java (Figure \ref{codefig}), along with the
corresponding Abstract Syntax Tree (AST, Figure \ref{astfig})
and the program function (Figure
\ref{progfunction}).  As for the Mathematica equation that
is generated from the AST, it is available online at\\ \tabeq
{\tt{\small http://web.njit.edu/\~{}mili/matheq.pdf}}.\\
The nesting structure of the equation reflects the structure of the
AST, which is itself a reflection of the level of nesting of the
program's control structure.  By solving this equation in the
primed variable ({\tt xP, yP, tP, iP, jP, kP}) as a function of
the unprimed variables ({\tt x, y, t, i, j, k}), we get the
function that the program defines from its initial states to its
final states.  Note that even though the program has three 
branches, its function has only two terms;  we will see in
section \ref{acvesect} why that is the case.
\begin{figure}
{\small{\bf\begin{verbatim}
public class Main 
   {
    public static int f(int x) 
       {x= 7*x+7; return x;}

    public static void main(String argv[]) 
       {int x, y, t, i, j, k;
        // read x, y, i, j, k, t;
        Label L1;
        t= i-j;
        j= i+5;
        if (i>j) 
           {x= 0;
            y= f(x);
            while (i!=j) 
                {i= i+k;
                 k= k+1;
                 i= i-k;
                 y= f(y);
                 Label L2;}
            Label L3;}
        else 
           {if (j>i) 
               {while (j != i) 
                   {j= j+k;
                    k= k-1;
                    j= j-k;
                    y= f(y);};
                Label L4;}
           else 
              {while (t!=i) 
                 {for (int z=0; z!=y; z=z+1) 
                     {x= x+1;}
                  y= x-y;
                  t= t+1;}
               Label L5;}
            }
        k= i+j;
        j= 2*k;
        Label L6;}
    }
\end{verbatim}}}
\caption{\label{codefig}Sample Java Code}
\end{figure}

%

\begin{table*}
\begin{center}
\begin{tabular}{|l|}
\hline
$\{\left(\langle\begin{array}{c}x\\y\\t\\i\\j\\k\end{array}\rangle,
\langle\begin{array}{c}xP\\yP\\tP\\iP\\jP\\kP\end{array}\rangle\right)|
(i<j) \wedge \left(\begin{array}{l} xP=x\wedge\\
yP=\frac{-7+7^{j-i}(7+6y)}{6}\wedge\\
tP=i-j\wedge\\
iP=i\wedge\\
jP=i\wedge\\
kP=k+j-i\end{array}\right)\}$\\
$\cup$\\
$\{\left(\langle\begin{array}{c}x\\y\\t\\i\\j\\k\end{array}\rangle,
\langle\begin{array}{c}xP\\yP\\tP\\iP\\jP\\kP\end{array}\rangle\right)|
(i=j\wedge j\geq 0) \wedge \left(\begin{array}{l}
xP=y Fib(j)+x Fib(j+1)\wedge\\
yP=y Fib(j-1)+x Fib(j)\wedge\\
tP=j\wedge\\
iP=i\wedge\\
jP=j\wedge\\
kP=k\end{array}\right)\},$\\
\hline
\end{tabular}
\caption{\label{progfunction}Program Function}
\end{center}
\end{table*}

\section{Querying Source Code at Scale}
\label{querysect}

\subsection{Assume, Capture, Verify, Establish}

Computing the function of a program is a valuable capability, but it
may be too much of a good thing:
\begin{itemize}
\item If the program being analyzed is too large, its function is
likely to be large and complicated as well.  A programmer who abhors
poring over pages of code, will probably not relish the prospect of
poring over pages of mathematical expressions instead.
\item While computing the function of a program in all its detail
is a commendable goal, it runs the risk that we capture secondary /
insignificant functional details
alongside the important functional attributes of the program.
\item If the function of the program differs from what the user
expects, it may be helpful for the user to analyze the program
at a smaller scale, to understand which part of the code
is faulty.
\end{itemize}
For all these reasons, we propose a set of functions that enable
a user to query the source code at specific locations (labels) in
the program, for specific sections of code.
If the programming language does not have provisions for labels, we 
create artificial labels by means of dummy object declarations, in
such a way that we have no impact on the program semantics.  Also,
we give the user the ability to highlight a part of the source code
and assign it a name, so that it can be the subject of subsequent
queries.
These
functions are invoked as part of an interactive session, and are answered
by the system on the basis of a static analysis of the source code
(see Figure \ref{acvefig}).
\begin{itemize}
\item {\em Assume()}. 
\begin{itemize}
\item {\em @L:  Assume(C)}, where 
$L$ is a label and $C$ is a unary predicate on the state 
of the program, formulates an assumption that the user makes about the
state of the program at label $L$.  In particular, this function can be
used to formulate the pre-specification of a program or a subprogram
or to serve as an assumption for a local proof.
\item {\em @P:  Assume(C)}, where $P$ is a
named program part and $C$ is a
binary predicate on the state of the program, formulates an assumption 
that the user makes about the function of $P$.  In particular, this
function can be used to formulate a hypothesis on a program part in
the process of proving a property about a larger encompassing 
program.
\end{itemize}
\item {\em Capture()}.
\begin{itemize}
\item {\em @L: Capture()}, where $L$ is a label, calls on the system
to generate two unary conditions: A {\em Reachability Condition},
which is the condition on the initial state of the program under which
execution reaches label $L$; and a {\em State Assertion}, which
captures everything that is known about the state of the program
at label $L$, considering the
assumptions that may have been recorded previously,
and in light of the code that is executed between the start of the
program and label $L$.
\item {\em @P: Capture()}, where $P$ is a program part, calls on
the system to generate a binary predicate in the state of the program
(in $(s,s')$),
which captures everything that is known about the function of $P$.
\end{itemize}
\item {\em Verify()}.
\begin{itemize}
\item {\em @L: Verify(C)}, where $L$ is a label and $C$ is a unary
condition on the state of the program, calls on the system to return
TRUE of FALSE depending on whether condition $C$ is assured to be
true at label $L$ or not.  For the sake of documentation, we may also
include the condition under which label $L$ is reachable
(if a label is unreachable, saying that some condition holds or
does not hold at that label may be misleading).  A user may invoke
{\em Verify(C)} to check a program's postcondition for correctness.
\item {\em @P:  Verify(C)}, where $P$ is a program part and $C$ is
a binary condition on the state of the program, calls on the system
to return TRUE of FALSE depending on whether program $P$ 
refines the binary relation defined by $C$.  A user may invoke
{\em Verify(C)} to check, for
example, whether a program or program part
satisfies some safety condition.
\end{itemize}
\item {\em Establish()}.
\begin{itemize}
\item {\em @L:  Establish(C)}, where $L$ is a label and $C$
is a unary condition that does not hold at label $L$.
The call to this function deploys program repair techniques 
\cite{gazzola2019} to 
generate mutants of the path from the first executable statement
of the program to label $L$, and selects a mutant that makes
condition $C$ true at label $L$
(and of course making label $L$ reachable as well).  
\item {\em @P:  Establish(C)}, where $P$ is a program part and
$C$ is a binary predicate on the state of the program that is
not refined by the function of $P$.  The call to this function
deploys program repair technology to generate mutants of $P$
and select a mutant that makes {\em @P: Verify(C)} return
TRUE.
\end{itemize}
\end{itemize}
We refer to these four functions as the {\em ACVE functions},
where ACVE is the acronym made up of the initials of these 
functions.  
These functions may sound similar to those of Dafny
\cite{leino2010,leino2023}, which features functions such
as {\em assert()}, {\em requires()}, {\em ensures()}.  Our
work is actually different from {\em Dafny} in fundamental
ways:
\begin{itemize}
\item The Dafny functions ({\em assert()}, {\em requires()},
{\em ensures()}) apply to programs written in the Dafny
programming language, whereas our ACVE functions can be
applied to any C-like programming language (provided we
transform its source code to an AST).
\item Dafny's functions are part of the Dafny programming
language, and have an impact on the execution of Dafny
programs, whereas the ACVE functions are used offline to
query the source code of programs, but have no impact on
the semantics of the programs.
\item Dafny's functions are a way for the programmer to
add redundancy to the program by mixing programming clauses
with specification clauses, whereas the ACVE functions are
used to derive the specification of the program as written.
\end{itemize}

\subsection{System Specification Through a Use Case}

\begin{figure*}
\begin{center}
\includegraphics[width=6in]{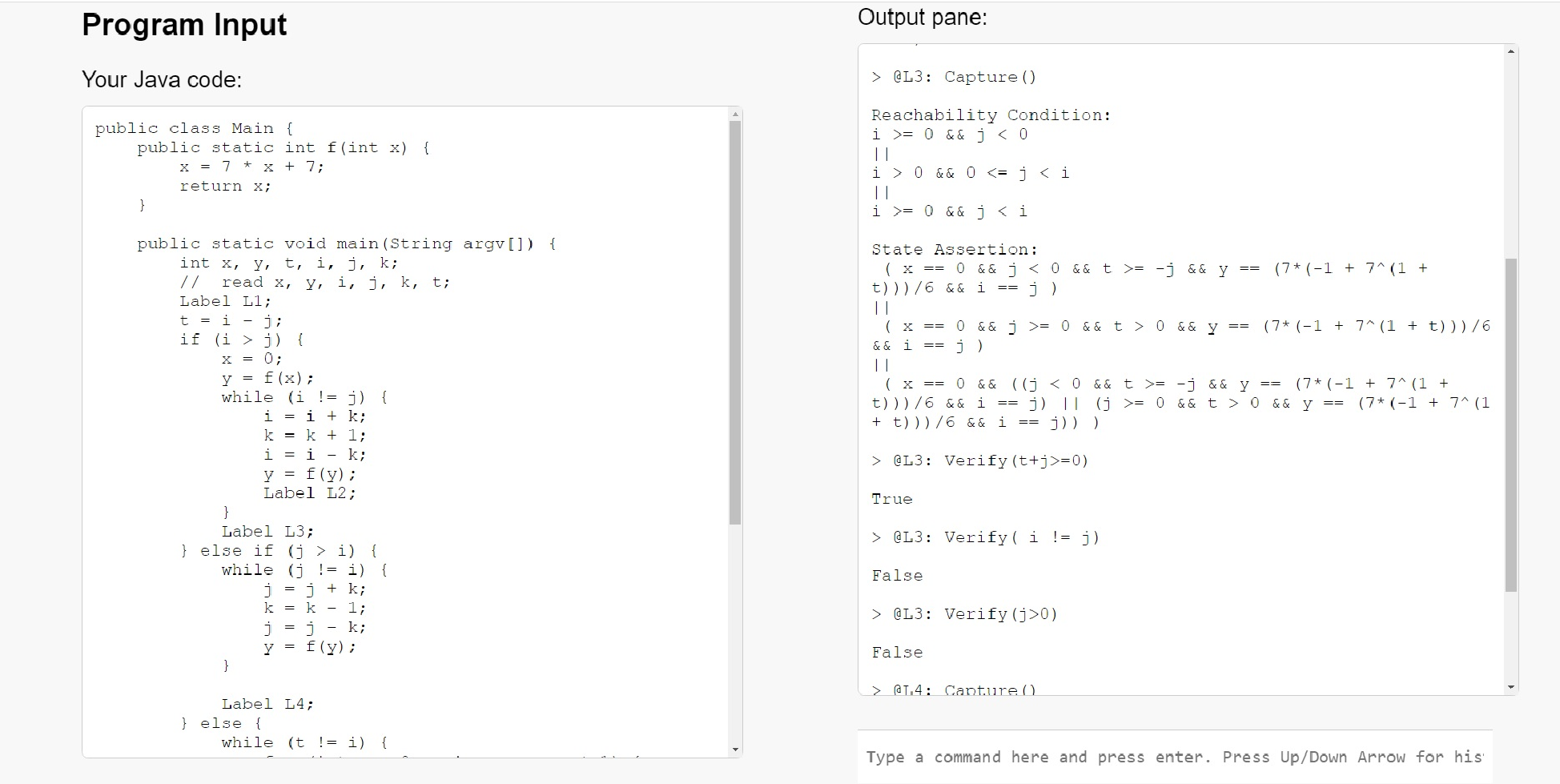}
\caption{\label{acvefig}Screenshot of a Session}
\end{center}
\end{figure*}

In its most basic form, the main GUI of ACVE
includes three windows (see Figure \ref{acvefig}):  A window that
contains the source code on the left side; a small window on the right
lower corner, where the user enters commands/ queries; and a larger 
window on the right, where ACVE echoes the queries and posts its
replies.  In this section we show a sample interactive session,
where the user uploads the Java program shown in Figure 
\ref{codefig}, then submits a sequence of queries.
Notice that this code contains a function definition, function
calls, branching logic, while loops, a for loop, and the for
loop is nested within the while loop; also it includes labels, which
we use to as reference points for our queries.  We acknowledge
that this program is not computing anything useful, but we use it
to showcase the capabilities of ACVE.

\subsubsection{@L1:  Assume($i>=0$)}

Label $L1$ is located at the first executable statement
of the program, right after the program's inputs are
read; hence one can interpret
{\tt @L1: Assume(i>=0)} as specifying the pre-condition
of this program.  

The system takes note of this assumption:{\small\begin{verbatim}
(i>=0) assumed at label L1.\end{verbatim}}
This assumption will be taken into account in any subsequent
query;  we can override an {\tt Assume()} statement by
calling another {\tt Assume()} statement with a 
different condition (assumption) at the same label;
in particular, we can cancel an {\tt Assume()} statement by
subsequently calling {\tt Assume(TRUE)} at the same label.

\subsubsection{@L2:  Capture()}

When we submit this query, we want to know:
\begin{itemize}
\item Under what condition on the initial values of the
program variables is label $L2$ reached?
\item When execution does reach label $L2$, what do we know
about the state of the program variables at that label?
\end{itemize}

The system replies: {\small\begin{verbatim}
Reachability Condition:
j>=0 && i>j  || j<0 && i>=0

State Assertion:
1+i==j+t && x==0 && y==56 && t>0 && (j>=0||j+t>=0)\end{verbatim}}

The reader can easily verify that these answers are correct:
\begin{itemize}
\item {\em Reachability Condition}:
Note that label $L2$ is inside the body of the loop
which is inside the then-branch of the if-then-else;  
execution will reach label $L2$ if and only if $(i>j)$
to enter the then-branch, then $(i\neq j)$ to enter the
loop; since neither $i$ nor $j$ are changed prior to these
two tests, the condition $(i>j)$ applies to the initial state.
This, in addition to the assumption $(i\geq 0)$ provided by
the {\em Assume()} command, yields the reachability condition.
\item {\em State Assertion}:
The condition $(x=0)$ stems
from the assignment {\tt \{x=0;\}}, the condition $y=56$ stems from
the double application of function $f$ to $x$, which is initialized
to 0 ($f(f(0))=56$; the condition $t>0$ stems from the fact
that $t$ was assigned the expression $(i-j)$ and $L2$ is reached
when $(i>j)$; the condition $(1+i=j+t)$ stems from the fact
that $t$ is assigned the expression $(i-j)$ prior to the
if-then-else statement, and $i$ is decremented by 1 in the
body of the loop, hence at $L2$ $t$ satisfies the condition
$(t=i+1-j)$, from which we infer $(1+i=t+j)$;  the clause
$(j+t>=0)$ in the last conjunct stems from the assignment
{\tt \{t=i-j\}}, which establishes the assertion $(i=t+j)$
and the assumption made at label $L1$ that $i$ is greater
than or equal to zero, hence so is $(t+j)$ at label $L2$
since neither $t$ nor $j$ is modified between $L1$ and $L2$;
as justification of the clause $j\geq 0$, consider that
$j$ is non negative, then the condition $(t+j\geq 0)$ holds
necessarily, since $t$ is known to be positive at label $L2$.
\end{itemize}

\subsubsection{@L3: Capture()}

Whereas label $L2$ is placed inside the loop body, and queries
the state of the program at the end of the first iteration, 
label $L3$ is located outside the loop, and represents the
state of the program when the loop has terminated.  If we want to
enquire about the state of the program at the exit of the loop,
and under what condition label $L3$ is reached at all,
we submit the query {\em @L3: Capture()}.  

The system replies: {\small\begin{verbatim}
Reachability Condition:
i>=0 && (j<0 || i>j)

State Assertion:
i==j && x==0 && 7^(2+t)==7+6*y && t>0  && (j>=0||j+t>=0)
\end{verbatim}}

We analyze this output to check for validity:
\begin{itemize}
\item {\em Reachability Condition}:  As a reminder,
this is a condition on the initial state under which
label $L3$ is reached.  The clause $(i\geq 0)$ stems
from the {\em Assume()} statement at label $L1$ and 
the clause $(j<0\lor i>j)$ stems from the condition
under which the first branch of the program is
taken, which is $(i>j)$:  if $j$ is negative then the
condition $(i>j)$ is satisfied vacuously since $(i\geq 0)$;
else the condition $(i>j)$ is imposed separately.

\item {\em State Assertion}:  The conditions $(x=0)$,
$(t>0)$ and $(j\geq 0\lor j+t\geq 0)$ are justified in the
same way as before, since $x$, $t$ and $j$ are not 
modified between label $L2$ and label $L3$ (only variables
$i$, $k$, and $y$ are, if the loop iterates more
than once).  Condition $(i=j)$ clearly holds
at $L3$ since $L3$ marks the exit from the loop, which
iterates as long as $(i\neq j)$.  As for the condition
$(7^{2+t}=7+6y)$, it can be justified as follows:
First note that by adding $k$ to $i$, incrementing it, then
subtracting it from $i$, the loop is decrementing $i$
by 1 at each iteration, all while not changing $j$;
also, the loop starts with $i$ greater than $j$, and exits
when $i$ equals $j$, which means the loop iterates $(i-j)$
times; but note that $t$ has been initiatlized with
$(i-j)$ prior to the loop, and $t$ has remained 
constant through execution of the loop, hence the value
of $t$ at label $L3$ 
measures the number of iterations of the loop (while
$(i-j)$ does not, since at $L3$ $(i-j)=0$).  The loop
starts with $y=f(0)=7$.  After one iteration, we get:\\
\tabeq $y=f(f(0))=7^2+7$.\\
After two iterations, we get:\\
\tabeq $y=f(f(f(0)))=7^3+7^2+7$.\\
Etc... after $t$ iterations, we get:\\
\tabeq $y=f^{t+1}(0)=7^{t+1}+7^t+...+7^3+7^2+7$.\\
If we factor out 7 from this expression, we find:\\
\tabeq $y=7\times (7^t+7^{t-1}+...+7^2+7+1)$.\\
The parenthesized 
expression is a geometric sum, whose closed form is known:\\
\tabeq $y=7\times\frac{7^{t+1}-1}{7-1}.$\\
From this expression we derive:\\
\tabeq $6y+7 = 7^{t+2}$.\\
\end{itemize}

\subsubsection{@L3:  Verify()}

If the user finds the output of {\em Capture()} too
detailed and just wants to check a simple minimal 
condition (e.g. a safety condition), then the user
may submit {\em Verify()} queries.  We give below
three simple examples at label $L3$, along with the
system's response. {\small\begin{verbatim}
@L3:  Verify(t+j>=0)
>> TRUE
@L3:  Verify(i!=j)
>> FALSE
@L3:  Verify(j>0)
>> FALSE
\end{verbatim}}

\subsubsection{@L4: Capture()}

Label $L4$ is located at the end of the second branch of the
nested if-then-else, right after the exit from the while loop.
We submit query {\em @L4: Capture()} to enquire on the
condition under which this label is reached, and the state
assertion that holds at this label.  The system replies:{\small
\begin{verbatim}
Reachability Condition:
>> FALSE

State Assertion:
>> FALSE
\end{verbatim}}

To understand this outcome, consider this:  The body of the
loop that precedes label $L4$ adds $k$ to $j$ at each iteration,
decrements $k$, then subtracts $k$ from $j$; so that each
iteration increments $j$ by 1; notice also that the loop does not
change $i$, and it is supposed to terminate when $j$ equals $i$.
But this branch of the if-then-else is entered when $j$ is greater
than $i$.  So the loop starts with $j$ greater than $i$ and makes
it greater at each iteration until it becomes equal to $i$, but
that will never happen, hence whenever this branch of the if-then-else
is entered, it leads to an infinite loop and label $L4$ is never
reached.  

\subsubsection{@L4:  Verify()}

Strictly speaking, a query such as:  {\em @L:  Verify(C)} means:
if execution reaches label $L$,
does condition $C$ hold for the program state at label $L$?
This means that if label $L$ is unreachable, then {\em @L: 
Verify(C)} ought to return TRUE for any $C$.  Indeed,{\small
\begin{verbatim}
@L4:  Verify(1==1)
>> TRUE
@L4:  Verify(1==0)
>> TRUE
\end{verbatim}}

Because some users may be confused or misled by this possibility,
we may add the reachability condition to the reply of {\em Verify()};
also, when the reachability condition is FALSE, we may want to
alert the user, rather than simply posting the state assertion
as TRUE.

\subsubsection{@L5:  Capture()}

Label $L5$ is located in the third branch of the nested if-then-else,
and it follows a nested loop.  We resolve to query the program
at this label, by means of {\em Capture()}.  

The system replies (where $\kw{Fib}$ is the
Fibonacci function):{\small\begin{verbatim}
Reachability Condition
   (i==j) && (j>=0)

State Assertion
   (i==t&&j==t&&(y==(x*Fib[t])/Fib[t+1]&&t>0)
   ||
   (t==0&&i==0&&j==0)
\end{verbatim}}

Justification:  
\begin{itemize}
\item {\em Reachability Condition}.
The third branch of the nested if-then-else is
entered only if $(i=j)$; since $t$ has been assigned the expression
$(i-j)$, we can infer that when this branch is entered, $t$ has value
zero.  In the loop, $t$ is incremented by 1 at each iteration, and the
loop ends when $(t=i)$, which means the only condition under which 
this loop terminates is that $i$ is non-negative.  
Hence label $L5$ is reached if and only if $(i=j) \wedge (i\geq 0)$.
\item {\em State Assertion}.  The state assertion includes two
disjuncts:  The case when the loop does not iterate at all is
captured by the condition $(t=0\wedge i=0\wedge j=0)$; indeed
$t=0$ is assured whenever execution enters this branch; if $i$
is also equal to zero, then the loop does not iterate, and since
$(i=j)$ in this branch, $j$ is also known to be equal to zero.

If the loop does iterate at least once then we know that 
when the loop terminates $t$ equals $i$, whence $t$ also equals $j$
since $(i=j)$; we also know that $t>0$ since $t$ is incremented by
1 at each iteration and the loop has iterated at least once. As
for the condition that $y$ equals
$$(x\times Fib(t))/Fib(t+1),$$
note that the for loop inside the while loop is just adding
$y$ to $x$, which means that the body of the while loop can be
rewritten as:\\
\tabeq {\tt\{x=x+y; y=x-y; t=t+1;\}}\\
This loop computes in $x$ and $y$ the following Fibonacci functions:\\
\tabeq $x'=x Fib(t+1)+y Fib(t)$\\
\tabeq $y'=x Fib(t)+y Fib(t-1)$.\\
The equation:\\
$$y=(x\times Fib(t))/Fib(t+1),$$
stems from the formulas of $x'$ and $y'$ and depends on properties of
the Fibonacci function; note that the division alluded to in this
equation is an integer division, i.e. $y$ is the floor of the expression
on the right.
\end{itemize}

\subsection{A Video Illustration}

A brief demo of these interactions is available in
{\tt mp4} format at: \\
\tabeq 
{\tt\small http://web.njit.edu/\~{}mili/acvedemo.mp4}.\\
The window that opens briefly following each query represents
the execution path that is generated for that query;
analysis of that execution path helps to compute the
ourput of the query.

\section{Using Path Functions to Answer Queries}
\label{acvesect}

Now that we have showcased the operation of our prototype,
and justified/ validated each answer that the prototype has
returned, we discuss, in the next sections, how we achieve this
functionality.

\subsection{Execution Paths}

The semantic rules discussed in section \ref{semanticssect} apply
to well formed synactic constructs of the programming language, but
with queries applying to random labels and random program parts, we
have to make provisions for analyzing arbitrary execution paths
through a program.  Also, when an {\em Assume()} command is submitted
at label $L$, we must reflect this command by recording at label $L$
a Java compatible statement that represents the assumed condition.

To this effect, we introduce three function declarations in Java:{\small
\begin{verbatim}
   public static void Assume(Boolean x) {}
   public static void TrueTest(Boolean x) {}
   public static void FalseTest(Boolean x) {}
\end{verbatim}}
and we use them to represent execution paths in Java-like syntax.
An execution path is a sequence of Java statements which may 
include, in addition to Java statements, any one of the three
statements above.
\begin{itemize}
\item {\tt Assume(C)} is inserted in any location where the user
has submitted an {\tt @L: Assume(C)} command.
\item {\tt TrueTest(C)} is inserted in any sequence of statements
between a statement that precedes a test (as in {\tt if (C) ..}
or {\tt while (C)..}) and the statement that follows it if the test
is TRUE.
\item {\tt FalseTest(C)} is inserted in any sequence of statements
between a statement that precedes a test (as in {\tt if (C) ..}
or {\tt while (C)..}) and the statement that follows it if the test
is FALSE.
\end{itemize}
Execution paths are generated from an analysis of the Abstract Syntax
Tree of the program, according to the following summary algorithm.  
\begin{enumerate}
\item We locate the label which is the target of the execution path.
\item We let {\em CurrentNode} be the AST node of the label.
\item \label{stepnumber} If {\em CurrentNode} is in a block,
then we discard all the nodes that follow it in the block.
\item If {\em CurrentNode} is in the body of a while loop
or in the if-branch of an if-then-else statement or an if-then
statement, we add {\tt TrueTest(C)} in its parent node in the AST,
prior to the link that points to it.
\item If {\em CurrentNode} is in the body of an else-branch 
of an if-then-else 
statement, we add {\tt FalseTest(C)} in its parent node in the AST,
prior to the link that points to it.
\item We Replace {\em CurrentNode} by its parent in the AST.
\item If we reach the node of {\tt main}, then 
goto step \ref{laststep}, else
goto step \ref{stepnumber}.
\item \label{laststep} Replace each label at which an {\em Assume()}
has been declared by the corresponding {\em Assume()} statement.
\item Convert the remaining AST onto text format;
that is the path from the start of execution to the indicated label.
\end{enumerate}

\subsection{Path Functions}

The function of a path is then computed exactly as we compute the
function of any Java program (discussed in section
\ref{semanticssect}, above), modulo the following semantic definitions:
\begin{itemize}
\item {\tt A2M(Assume(C))}\\
\tabeq \tabeq {\tt = C \&\& x1==x1P \&\& ... \&\& xn==xnP.}
\item {\tt A2M(TrueTest(C))}\\
\tabeq \tabeq {\tt = C \&\& x1==x1P \&\& ... \&\& xn==xnP.}
\item {\tt A2M(FalseTest(C))}\\
\tabeq \tabeq {\tt = !C \&\& x1==x1P \&\& ... \&\& xn==xnP.}
\end{itemize}
As an illustrative example, we show below the paths that ACVE
generated for the queries at labels $L2$ and $L3$, and their
corresponding path functions.

For L1-L2:{\small\begin{verbatim}
Assume(i>=0);
t=i-j;
TrueTest(i>j);
x=0; y=f(x);
TrueTest(i!=j);
i=i+k; 
k=k+1; 
i=i-k; 
y=f(y);
\end{verbatim}}
The function of this path is:{\small\begin{verbatim}
iP=i-1&&tP=i-j&&jP==j&&kP==k+1&&xP==0&&yP==56
      &&(i>=0||j>=0) &&(j<0||j<i)\end{verbatim}}

For L1-L3:{\small\begin{verbatim}
Assume(i>=0);
t=i-j;
TrueTest(i>j);
x=0; y=f(x);
while (i!=j)
   {i=i+k; 
    k=k+1; 
    i=i-k; 
    y=f(y);}\end{verbatim}}
Note that in this path the loop appears in full.
The function of this path is:{\small\begin{verbatim}
tP=i-j&&iP=j&&jP==j&&kP=i+k-j&&xP==0
      &&yP==(7^(2+i-j)-7)/6
      &&(i>=0||j>=0) &&(j<0||j<i)\end{verbatim}}

\subsection{Query Semantics}

In the previous section we discussed how to generate the 
execution path leading
to a label, and how to compute the function of the path; in
this section we discuss how to use the generated function to
answer {\em Capture()} and {\em Verify()} queries.
\begin{itemize}
\item
{\em @L:  Capture()}.  Let $P$ be the function of the path defined
by label $L$.  The reachability condition of $L$ is simply the
domain of $P$; we formulate it as follows:\\
\tabeq $RC\equiv \exists sP:  (s,sP)\in P$.\\
As for the state assertion at label $L$, it is the range of
function $P$; but remember, we must formulate it as a condition
on the state of the program at label $L$, hence we cannot simply
say $(\exists s: (s,sP)\in P)$, since we must formulate this
condition in $s$, not in $sP$.  Whence the following formula:\\
\tabeq $SA\equiv \kw{sP2s}(\exists s: (s,sP)\in P)$,\\
where $\kw{sP2s}$ is defined as follows:\\
\tabeq $\kw{sP2s}(F(sP))=(\exists sP: s==sP\wedge F(sP))$.
\item
{\em @L:  Verify(C)}.  In principle, this query should simply
return the result of proving the theorem:\\
\tabeq $SA\Rightarrow C$,\\
where $SA$ is the state assertion derived from {\em @L: Capture()}.
But it is possible to also add the reachability condition of label
$L$, for the benefit of users; also, when the reachability condition
is actually FALSE, we may warn the user rather than return TRUE
vacuously.
\end{itemize}

\section{Conclusion}
\label{concsect}

\subsection{Summary}

In this paper we discuss the design and preliminary implementation
of a tool that enables the user to query a program by making assumptions
about program states and program parts and by asking the system to
reveal semantic properties that hold at program labels or 
about program parts.  Specifically, we propose four functions:
\begin{itemize}
\item {\em Assume()}, which enables the user to make assumptions
about program states or program parts.
\item {\em Capture()}, which enables the user to elicit semantic
information about program states or program parts.
\item {\em Verify()}, which enables the user to confirm or disprove
assumptions the user makes about program states or program parts.
\item {\em Establish()}, which enables the user to modify a program
to make it satisfy a tentative assumption that was disproven.
\end{itemize}
As of the due date, we have implemented the first three functions and
are implementing the fourth; the fourth function uses existing
program repair
freeware to generate mutants of the original program so as to 
satisfy a {\em Verify()} condition; unlike traditional program
repair technology, it does not use testing to select repair candidates,
but checks correctness with respect to precisely formulated 
specifications (given as the argument of the
{\tt Establish()} command)
and is based on a static analysis of generated repair
candidates.

\subsection{Critique and Threat to Validity}

The proposed approach is based on our ability to compute or
approximate the function of a program, on the basis of a static
analysis of its source code.  This includes programs that have 
loops, nested (in principle) to an arbitrary degree.  The analysis
of loops relies on the availability of a database of recognizers,
and can only work to the extent that it can match the program
against its recognizers.  This is clearly a serious limitation,
which can only be addressed if we plan to derive and store a
large database of recognizers; but a large database imposes
a heavy penalty to match the source code against all possible
recognizers.

\subsection{Related Work}

Programs are detailed, information-rich, potentially
complex artifacts, hence
analyzing programs is typically a tedious, non-trivial, error-prone 
exercise.  It is natural in such circumstances to seek the
support of automated tools, but these come up against the
difficulty to capture the semantics of iterative statements.
Capturing the semantics of iterative statements amounts to
reverse engineering the inductive argument that was used in 
their design, a daunting task that requires a great deal of
creativity, and is very difficult to automate.
It is difficult to synthesize several decades of research in one
paragraph, so our attempt is bound to indulge in gross oversimplification;
with this caveat in mind, we can discern two broad approaches to the
analysis of iterative programs:  
\begin{itemize}
\item
The first is to approximate the
function of a loop by unrolling it a 
small (user-specified, or fixed by default) number of times
\cite{dockins2016,frenkeletal2020,santos2020,cadar2013,torlak2014}.
\item The second is to capture the semantics of a loop
by attempting to derive its loop invariant
with respect to a specification in the form of a
(precondition,postcondition) pair
\cite{gupta2009,colon2003other,ernst2007,ernst2001,kroening2010other,%
ancourt2010,mcmillan2008,PodelskiR11,PodelskiR04,kovacs2007full,%
henzinger2008full,kovacs2009full,cousot1977full,meyer2010full,%
mclean2010full,zuleger2010full,gulwani2010,iosif2010full,fubastani2008full,%
lahiri2004,rodriguezcarbonnell2004full,colon2003,sankaranarayanan2004,
stark98}.
\end{itemize}
The first approach was always understood to be a compromise,
adopted in the absence of better or more complete solutions.
Despite some successes,
the promise of the second approach remains for the most part
unfulfilled, for several reasons:
\begin{itemize}
\item First, because the generation of a loop invariant
is very complex, as it depends on three parameters: the
loop's pre-specification, the loop's post-specification, and its
guarded loop body.  
\item Second, the loop invariant must satisfy three
conditions, and appears on the left side of implications
($\Rightarrow$) in some and on the right side of
implications in others.  To make the loop invariant
strong enough for the conditions where it is on the left-hand
side of implications and weak enough for the conditions
where it is on the right-hand side is a delicate balancing
act that requires much analysis and creativity.
\item Third, before we can initiate the search of a loop
invariant, we must have a specification of the loop in the
form of a (precondition, postcondition) pair.  
This is something of a chicken-and-egg situation because
to ensure that the loop is correct with respect to
a (precondition, postcondition) we need to generate a loop
invariant, but to generate the loop invariant we need to
have a (precondition, postcondition) specification.
\item Fourth, if the generation for a loop invariant fails, we
have no simple way to tell whether it is because our generation
algorithm is flawed or because the while loop is not correct
with respect to the specification. 
\item Fifth, to the best of our knowledge, no method or
tool can generate loop invariants for nested loops.
\end{itemize}
In this paper we analyze loops by means of invariant relations,
which are binary relations that hold between states of the
program that are separated by an arbitrary number of iterations.
Using invariant relations rather than loop invariants
(aka {\em invariant assertions}), we address some of the
difficulties cited above:
\begin{itemize}
\item An invariant relation depends exclusively on the while
loop, regardless of what specification the loop is supposed
to satisfy.
By virtue of this property, invariant relations bear some
resemblance to the {\em inductive assertions} used, e.g. in
IC3 \cite{hassan2013}; but the generation of invariant
relations, which we discuss in this paper, is very different
from the generation of inductive assertions in IC3.
\item Invariant relations reveal the function of the loop 
{\em as written}, hence do not require that we make guesses/
assumptions as to their specification.
\item Since invariant relations can be used to compute the
function of a loop, nested loops can be handled by replacing
an inner loop by its function and proceeding as if it were
an assigmnent.
\end{itemize}
Recent work on program summarization attempts to approximate
the function of a program using machine learning methods
or recurrent neural nets
\cite{choi2020source,zhou2020effective}; our work is
different since it aims to compute the exact program function,
in all its detail, and uses a systematic analysis of the source
code to this effect.
In \cite{chen1998c++} Chen et al present a tool that analyzes
C++ code to compute executions paths, and determine which
classes or functions are called for a given input.
In \cite{baudin2021}, Baudin et al. discuss the design,
implementation and evolution of a sprawling environment they
have been developing to perform the static analysis of C
code;  {\em Frama-C} is a collaborative, open-source platform
for code analysis and verification, based on formal methods, 
notably on the the theory of abstract interpretation
\cite{cousot2008}; being open source, {\em Frama-C} makes
provisions for plug-ins, and uses these to enrich the tool's
functionality in a modular fashion.  Because it is based on
abstract interpretation, {\em Frama-C} is optimized to detect
run-time errors, just like {\em Astree}
\cite{kastner2010}, which shares the
same theoretical pedigree.  In our approach, we aim to capture the
full function of a program; to the extent that we can do so while
taking into account computer limitations, run-time errors are instances
where the input is outside the domain of the program's function.

\subsection{Prospects}

Our short term plan is to implement, test and fine-tune the
program repair capability that enables to support the {\em
Establish()} function; this would differ from current
program repair practice \cite{gazzola2019} in the sense
that we do not need test data (and the dilemmas that
come with selecting test suites), and that repair candidates
are selected on the basis of satisfying a {\tt Verify()}
clause (rather than running successfully on the
selected test suite).  

Our medium term goal is to investigate the automated derivation
of recognizers.  Mathematica has an interesting capability that
we find useful in this regard: function {\tt RSolve()}, which
can be used to derive closed form expressions from recurrence
equations.  Indeed, if we formulate the $F$ term of a recognizer
(re:  Table \ref{rectab}) as a recurrence formula, then we
quantify the recurrence variable away ($\exists n:$), we obtain
an invariant relation for $F$.  For example, if we apply {\tt RSolve()}
to the recurrence equations:
\begin{verbatim}
x[n+1]=x[n]+1 && y[n+1]=y[n]-1
\end{verbatim}
we obtain the following closed form:
\begin{verbatim}
x[n]=x0+n && y[n]=y0-n.
\end{verbatim}
Applying the existential quantifier to this formula ($\exists n$) and
renaming {\tt x0, y0, x[n], y[n]} as respectively 
{\tt x, y, xP, yP}, we find:  {\tt x+y=xP+yP},
which is the correct invariant relation. The same approach has
enabled us to derive recognizers that would have been nearly impossible
to generate by hand.  But this method is limited 
exclusively to recognizers
that deal with numeric data types, and that {\tt RSolve()} is able to
solve.

\subsection*{Acknowledgement}

This research is partially supported by NSF under grant number
DGE2043204.  The authors are very grateful to Dr Jules Desharnais,
Laval University, for his insights on the sufficiency condition
of Proposition \ref{sufficiencyprop}.

\end{document}